\documentclass[lettersize,journal]{IEEEtran}
\usepackage{amsmath,amsfonts}
\usepackage{array}
\usepackage[caption=false,font=normalsize,labelfont=sf,textfont=sf]{subfig}
\usepackage{textcomp}
\usepackage{stfloats}
\usepackage{url}
\usepackage{verbatim}
\usepackage{graphicx}
\usepackage{cite}
\usepackage{color}
\usepackage{multirow}
\usepackage{makecell}
\usepackage{empheq}

\usepackage{threeparttable}
\usepackage{balance}
\usepackage{diagbox}
\usepackage{algorithmic}
\usepackage{cite} 
\usepackage{amsmath,amssymb,amsfonts}
\usepackage{algorithmic}
\usepackage{graphicx}
\usepackage{textcomp,lineno,hyperref,chngcntr,caption,wasysym,textcomp,mathrsfs,indentfirst,subfig,verbatim,url} 
\usepackage{verbatim}
\usepackage{algorithm}
\usepackage{algorithmic}
\usepackage{diagbox}
\makeatletter
\def\test@relax{\relax}
\let\save@fnum@lstlisting\fnum@lstlisting
\def\fnum@lstlisting{%
    \save@fnum@lstlisting
    \ifx\lst@caption\test@relax\expandafter\@gobble\fi
}
\makeatother

\begin{document}

\title{  Membership Reference Attack against Laplace Mechanism of Differential Privacy}
\author{Wen Huang, Zhishuo Zhang, Weixin Zhao, Jian Peng, Yongjian Liao, Yuyu Wang
%\author{IEEE Publication Technology,~\IEEEmembership{Staff,~IEEE,}
        % <-this % stops a space
\thanks{ Wen Huang, Weixin Zhao and Jian Peng are with Sichuan University, Chengdu 610065, China.
    Zhishuo Zhang, Yongjian Liao, and Yuyu Wang are with the School of Information and Software Engineering, University of Electronic Science
    and Technology of China (UESTC), Chengdu 610054, China.
    }% <-this % stops a space
\thanks{Manuscript received June 1, 2024; revised July 1, 2024.}}

\markboth{Journal of \LaTeX\ Class Files,~Vol.~1, No.~2, December~2023}%
{Shell \MakeLowercase{\textit{et al.}}: A Sample Article Using IEEEtran.cls for IEEE Journals}

%\IEEEpubid{0000--0000~\copyright~2023 IEEE}

\maketitle

\begin{abstract}
	The differential privacy is a widely accepted conception of privacy protection and the Laplace mechanism is a famous instance of differential privacy mechanisms to deal with numerical data. In this paper, we point out that the differential privacy does not take liner property of queries into account, resulting in information leakage. In order to show the information leakage, we construct a membership reference attacks against the Laplace mechanism. Concretely, we propose a method to obtain multiple independent identical distribution samples of linear query's answer under constrains of the Laplace mechanism. The proposed method is based on linear property of linear query and some background knowledge. Based on obtained samples, a hypothesis test method is used to determine whether a targert record is in data set.
\end{abstract}


\section{Introduction}
\par The differential privacy is state of the art conception for privacy protection because it captures a strong privacy guarantee. That is, even if adversaries know all records except one, adversaries cannot get any information of the record which adversaries don't know. The guarantee is an attractive property for preserving privacy analysis mechanisms and is a foundation of preserving privacy analysis tasks.

\par The Laplace mechanism is a famous instance to realize the idea of differential privacy by perturbation. The key idea of perturbation mechanisms is to mask  mechanism outputs' difference which is cased by presence or absence of one record with noise's fluctuation. When fluctuation is big enough, it is hard to determine whether the difference is cased by presence or absence of a record. In differential privacy, the biggest difference cased by presence or absence of one record is captured by conception of sensitivity denoted by $\Delta D$ and in general the noise's fluctuation is captured by conception of variance of noise distribution. As respect to the Laplace mechanism, the variance of noise distribution is $2(\Delta D/\epsilon)^2$(the $\epsilon$ is privacy budget). Intuitively speaking, in order to keep security of the Laplace mechanism, the sensitivity needs to be far less than variance, namely $\Delta D \ll 2(\Delta D/\epsilon)^2$. However, we find that for some special queries the inequality doesn't hold. For example, when the query function is counting and the privacy budget $\epsilon$ is 1, the $\Delta D=1$ and $2(\Delta D/\epsilon)^2 =2$.      

\par Based on the above observation, we thought that for some special queries noise added by the Laplace mechanism may be not big enough to mask the outputs' difference cased by presence or absence of a record. When we try to find a method to obtain multiple i.i.d.(independent identical distribution) samples for statistically analyzing, we find out another vulnerability of the differential privacy. That is, the differential privacy does not take liner property of queries into account, resulting in that it always can get multiple i.i.d. samples for a linear query. In order to demonstrate the information leakage of differential privacy mechanisms, we take advantage of vulnerabilities to construct a membership reference attacks method against to the Laplace mechanism. Our main contributions are as follows  

\par (1) We firstly point out that liner property of queries is a source of information leakage. To the best of knowledge, this paper is the first one to discuss information leakage cased by queries' property, namely linear property. 

\par (2) A method is proposed to obtain multiple i.i.d. samples for linear query's answer under constrains of differential privacy. In general cases, multiple i.i.d. samples for query's answer cannot be obtained directly. However, the linear query has two special properties. Firstly, the linear query has linear property. Secondly, global sensitivity of linear queries can be calculated easily. Based on the two properties, we propose a method to obtain multiple i.i.d. samples. 

\par (3) A membership reference attacks method against the Laplace mechanism is proposed. The proposed method is more general than other methods. Firstly, the proposed method can work for every records. The method which takes advantage of correlation of records only works for records which are related with each other. Secondly, the proposed method can work for every data set. The method which take advantage of inappropriate extension of sensitivity conception just work for partial data set. 

\subsection{Related Work}

\par The differential privacy is a very popular conception for privacy protection because of its good mathematics foundation. The conception of differential privacy is proposed by Dwork et al. \cite{6}. After the differential privacy is proposed, there are some analyses about its information leakage. The first angle is related to assumption of data generation. In differential privacy, records in data sets are assumed to be independent. But the assumption is not always true in some real application scenarios. For example, Bob or one of his 9 immediate family members may have contracted a highly infectious disease. The entire family would have been infected. An attacker asks the query “how many in Bob’s family have this disease” to infer if Bob has been infected. The true answer is of high probability to be either 0 or 10. Suppose the noisy answer returned is 12. If this answer is obtained by adding Laplace noise calibrated by the specified privacy parameter $\epsilon$, the attacker learns that the probability of 10 being the true answer is $e^{10\epsilon}$ times larger than the probability of 0 being the true answer, violating the expected privacy guarantee\cite{17}. There are other papers which research the correlation problem. Fredrikson et al. research the problem from genetic data\cite{18}. Wu et al. research the problem from game theory \cite{19}. The method to fix correlation problem is to find a quantity to quantify correlation and calibrate noise added by the quantity. Yang et al. quantify degree of correlation by Gaussian correlation model\cite{20}. Zhu et al. propose conception of correlated sensitivity to quantify degree of correlation\cite{24}. The strengths of the angel is that it is not related to certain mechanisms and the weakness of the angle is that method based on correlation is effective for correlated records not for every record. 

\par The second angle to analyze information leakage of differential privacy mechanisms is the magnitude of noise. There is a difficult trade-off in differential privacy, namely trade-off between magnitude of noise and data utility. In order to reduce the magnitude of noise, various conceptions are proposed such as elastic sensitivity \cite{22}. Some of these proposed conceptions of sensitivity are weak and lead to information leakage. Local sensitivity is example\cite{23}. The goal of local sensitivity is to release data with dataset-based additive noise. That is, the noise magnitude is determined not only by the queries which is to be released, but also by the dataset itself. But the noise calibrated by local sensitivity is too small, resulting in information leakage. For example, let $f_{med}(x) = median(x_1,x_2,\dots,x_n)$, where the $x_i$ is sorted real number from bound interval such as $[0,\Lambda]$. If the $f_{med}(x)$ is released with noise magnitude proportional to local sensitivity, some information will be leaked. Concretely, if the noise magnitude is proportional to local sensitivity,then the probability of receiving a non-zero answer  when $x_1=\dots = x_{m+1}=0$,$x_{m+2}=\dots=x_n=\Lambda$ is zero whereas the probability of receiving a non-zero answer when $x_1=\dots = x_{m}=0$,$x_{m+1}=\dots=x_n=\Lambda$ is nonnegligible. Where $n$ is odd and $m=\frac{n+1}{2}$. The analysis from extension of sensitivity conception is tricky because how big the noise magnitude needs to be is hard to quantify.

\par Different from the mentioned angles, we will discuss information leakage from another new angle. The first angle is about the special records which are correlated with each other and the second angle is about trade-off between noise magnitude and data utility. In this paper, from the view of query function, we show that differential privacy mechanism does not take liner property of query function into account, resulting in information leakage.

\subsection{Organization}
\par In the next section, background knowledge will be introduced briefly, including differential privacy, linear property, hypothesis test and model of membership reference attacks. In section 3, the membership reference attacks method will be presented. Then, a security analysis of Laplace mechanism and an instance of attack will be given in section 4. In section 5, results of experiments will be showed. At last, the conclusion will be claimed.

\section{background knowledge}  
%\par introduction, definition, local sensitivity, laplace , exponential
\par The introduction of background knowledge is divided into four parts, namely basic conceptions of differential privacy, linear property, hypothesis test and model of membership reference attacks.
\subsection{Differential Privacy} 
\par Differential privacy is the most popular conception of privacy protection. It formalizes a strong privacy guarantee: if one person's privacy is violated because he contributes his privacy information to a data set, the person's privacy is also violated even if he dose not contribute his data to the data set. So, differential privacy mechanism guarantees that its output is insensitive to any particular data record. That is, presence or absence of one record has limited influence on output of differential privacy mechanism. The presence or absence of one record in a data set is captured by conception of neighboring databases. Specifically, database $D$ and $D'$ are neighboring databases denoted by $D \sim D'$ if there is only one different record between $D$ and $D'$ denoted by $d(D,D')=1$. 
\par \textbf{Definition 1}(Differential privacy) Any random mechanism $M:D^n\to R^d$ preserves $\epsilon$-differential privacy if for any neighboring databases $D,D'$ such that $d(D,D')=1$ and for all sets of possible output $S$:
\begin{eqnarray*}
    P\{M(D)\in S\} \le e^\epsilon P\{M(D')\in S\}
\end{eqnarray*}
\par Here, the $\epsilon$ is the privacy budget. Privacy guarantee is quantified by the privacy budget. The smaller the privacy budget is, the stronger the privacy guarantee is.
%\par The security of differential privacy is based on a conception called $\epsilon$-indistinguishable. The conception is introduced by Dwork et al. in \cite{6}. Its formal definition is follows:
%\par \textbf{Definition 2}($\epsilon$-indistinguishable against k records) A mechanism is $\epsilon$-indistinguishabl if all pairs $y,y'\in D^n$ which differ in only $k$ entries, for all adversaries $A$, and for all transcripts $t$: 
%\begin{eqnarray*}
%	|ln \frac{P\{T_A(y)=t\}}{P\{T_A(y')=t\}} | < \epsilon
%\end{eqnarray*}

\par Laplace mechanism is a famous and foundational mechanism in the field of differential privacy. It is to deal with numerical data. The Laplace mechanism is based on a conception of global sensitivity. The global sensitivity quantifies the query's max difference which is cased by presence or absence of one record.

\par \textbf{Definition 2}(global sensitivity) For given database $D$ and query $q$, the global sensitivity denoted by $\Delta D$ is	
\begin{eqnarray*}
    \Delta D = \max \limits_{D \sim D'} |q(D)-q(D')|
\end{eqnarray*} 

\par The Laplace mechanism covers output's difference by noise and the noise is drawn from Laplace distribution with location parameter $\mu=0$ and scale parameter $b=\frac{\Delta D}{\epsilon}$.% The difference is cased by presence or absence of one record and the presence or absence of one record is captured by conception of neighboring databases as mentioned before.

\par \textbf{Definition 3} (Laplace mechanism) For given database $D$, query $q$ and privacy budget $\epsilon$, output of Laplace mechanism is 
\begin{eqnarray*}
    M(q,D,\epsilon) = q(D) + Lap(0,\frac{\Delta D}{\epsilon}) 
\end{eqnarray*} 
\par Here, the $Lap(0,\frac{\Delta D}{\epsilon})$ represents noise drawn from Laplace distribution with location parameter 0 and scale parameter $\frac{\Delta D}{\epsilon}$. 
%For simplicity of notation, $M(q,D)$ is used to present $M(q,D,\epsilon)$ in the following sections.  
\begin{comment}
    \par As for exponential mechanism, it is based on a function called utility function. The utility function denoted by $q$ is a measure for the quality of output. The inputs of utility function are database $D$ and the possible output r. The utility function should be insensitive for present or absence of any one record. The sensitivity of $q$ is $\Delta q$
    \begin{eqnarray*}
        \Delta q = \max \limits_{\forall r,D, D'} |q(D,r)-q(D',r)|  
    \end{eqnarray*} 
    
    \par The exponential mechanism is suitable for application scenarios where outputs are not real number or make no sense after adding noise. The exponential mechanism assigns  greatest probability to result whose value of utility function is greatest.
    
    \par \textbf{Definition 4} (Exponential mechanism) For a utility function $q$ and database $D$, the mechanism $M$ 
    \begin{eqnarray*}
        M(D,q) = \{return\ r\ with\ probability \propto exp^{\frac{\epsilon q(D,r)}{2\Delta q}}\} 
    \end{eqnarray*}   
\end{comment}

\par The differential privacy has many good properties which make it possible to build complex differential privacy mechanism by basic block algorithms. The two of them are Sequential Composition Theorem and Parallel Composition Theorem.       
\par \textbf{Sequential Composition Theorem} Let $A_1,A_2, \cdots, A_k$ be $k$ algorithms that satisfy $\epsilon_1$-DP,$\epsilon_2$-DP, $\cdots$,$\epsilon_k$-DP respectively. Publishing $t = <t_1,t_2,\cdots, t_k>$ satisfies $(\sum_{i=1}^{k}\epsilon_i)$-DP. Where, $t_1=A_1(D),t_2=A_2(t_1,D), t_3=A_3(<t_1,t_2,D>),\cdots, t_k=A_k(<t_1,t_2,t_3,\cdots,t_{k-1}>,D)$.   

\par \textbf{Parallel Composition Theorem} Let $A_1,A_2, \cdots, A_k$ be $k$ algorithms that satisfy $\epsilon_1$-DP,$\epsilon_2$-DP, $\dots$,$\epsilon_k$-DP respectively. Publishing $t = <t_1,t_2,\cdots, t_k>$ satisfies $(max_{i\in [1,2\cdots k]}\epsilon_i)$-DP. Where, $t_1=A_1(D_1),t_2=A_2(t_1,D_2), t_3=A_3(<t_1,t_2,D_3>),\cdots, t_k=A_k(<t_1,t_2,t_3,\cdots,t_{k-1}>,D_k)$ and $D_i \cap D_j = \emptyset $ for $i\ne j$,    

\subsection{Linear Property}
\par The linear query is one kind of foundational query. It is used widely. For example, sum and counting query are popular queries. The linear query has a very good property, namely linear property.
\par \textbf{Definition 4}(linear property) For linear query $q$ and data set $D_q$ such that $D_q = D_1 \cup D_2$ and $ \varnothing = D_1 \cap D_2 $, we have  
\begin{eqnarray*}
    q(D_q) = q(D_1) +q(D_2) 
\end{eqnarray*}
\par Counting query is one kind of linear query so it satisfies the linear property.  
%Counting function is one kind of query whose sensitivity is 1. And difference cased by presence or absence of a record is also 1. Because of these two factors, the proposed method is based on counting query.

\subsection{Hypothesis Test}
\par The hypothesis test is a statistic tool to judge whether fluctuation of experiment results is just due to randomness. Its main purpose is to guarantee confidence level of experiment conclusion. To that end, there are three steps in hypothesis test logically. Firstly, an assumption of experiment result's conclusion is made. The assumption needs to be contrary to conclusion which is achieved by experiment results. The assumption is called null hypothesis and the null hypothesis is usually denoted by $H_0$. The opposite hypothesis of null hypothesis is called alternative hypothesis and it is usually denoted by $H_1$. Secondly, a statistics is chosen. The statistics needs to be related to assumption and its distribution is known. Thirdly, the probability of experiment results is calculated through the known distribution. If the probability is smaller than a predetermined threshold, experiment conclusion is reliable. Because the procedure guarantees that when assumption which is contrary to experiment's conclusion is right, the experiment results occur with extremely small probability. That is, the experiment results are not due to randomness.    

\par The concrete steps of hypothesis test are clear. Firstly, the predetermined probability threshold needs to be chosen and it is called significance level $\alpha$. Secondly, the experiment results' probability is calculated and the probability is called $p$ value.  At last, a decision about hypothesis is made. Concretely, if the $p$ value is greater than the predetermined significance level $\alpha$, null hypothesis is accepted. And if the $p$ value is less than the predetermined value $\alpha$, the alternative hypothesis is accepted.

\par The one-sample t-test is useful hypothesis test method. It is used to judge whether samples come from a distribution with certain mean $\mu$. Here, the distribution mean is the mean of all possible experiment results and sample mean is the mean of experiment results which have already been obtained. Statistics $T$ is the statistic used in one-sample t-test and the form is as follows
\begin{eqnarray*}
    T &=& \frac{\bar{X} - \mu}{\frac{S}{\sqrt{m-1}}} \\
\end{eqnarray*}
\par Where, the $\mu$ is distribution mean. The $m$ is number of samples. When the samples are $\hat{a}_1,\hat{a}_2,\dots,\hat{a}_m$ and $\bar{X}$ is sample mean, we have
\begin{eqnarray*}
    \bar{X} = \frac{\hat{a}_1+\hat{a}_2+\dots+\hat{a}_m}{m} \\
\end{eqnarray*}
\par The $S$ is sample standard variance, 
\begin{eqnarray*}
    S=\sqrt{\frac{(\hat{a}_1-\bar{X})^2+(\hat{a}_2-\bar{X})^2+\dots+(\hat{a}_m-\bar{X})^2}{m}}   \\
\end{eqnarray*}   
\par There is a table where $p$ value can be found by the freedom and the value of $T$. The freedom is $m-1$ and the value of $T$ can be calculated as above.

\subsection{Membership Reference Attacks model }
\par Membership reference attack is to judge whether a record is in data set. Concretely, there is a data set $D$ with sensitive information and differential privacy mechanism $M$ is used to protect the data set $D$. An attacker has some background knowledge and the background knowledge is a data set $D_{know}$ which is a subset of $D$. The attacker has black-box access to differential privacy mechanism $M$. Given a data record $x$ which is not in $D_{know}$, the attacker guesses: whether the record $x$ is in data set $D$. The formal definition of membership reference attack is as follows:

\par \textbf{Definition 5}(membership reference attacks): Given a subset $D_{know}$ of data set $D$ , black-box access to Laplace mechanism $M$ and a record $x$, the attacker guesses whether the record $x$ is in the data set $D$ or not. 

$$ A(D_{know},x,M)=\left\{
\begin{array}{rcl}
    1      &      & {x\in D}\\
    0     &      & {x\notin D}
\end{array} \right. $$

\par Here, the $A$ presents membership reference attacks method. If $A$ asserts that $x$ is in data set $D$, it outputs 1. If $A$ asserts that $x$ is not in data set $D$, it outputs 0. 

\par There are there assumptions about the black-box access of differential privacy mechanism. 
\par (1) For same query, the black-box access returns the same answer no matter how many times the query is issued. This assumption is reasonable because if the attacker can get a new answer for same query every time the attacker issues the query, then the attacker can get multiple i.i.d. samples of the query's answer. With multiple i.i.d. samples, it is trivial for the attacker to inference information of records.

\par (2) For every different query, the black-box access returns an answer. The second assumption guarantees the utility of differential privacy mechanism. If the  black-box access refuses to answer queries which are issued first time, the utility of differential privacy is damaged, resulting in an useless mechanism.

\par (3) There is a threshold of privacy budget and when privacy budget consumed by answered queries is greater than the threshold, the black-box access aborts. The consumed privacy budget is accumulative and when consumed privacy budget is greater than certain limitation, it is trivial to lead information leakage.

\par The membership reference attacks method is effective when its success rate is greater than 50\%. The success rate of random guess is 50\% because there are only two possibilities. So, for any method which is effective, its success rate needs to be greater than success rate of random guess. 

\par The ability of attackers is limited. Firstly, the black-box access means that an attacker submits a query to Laplace mechanism and gets a result as return. The attacker cannot get any other information from the Laplace mechanism. Secondly, the attacker has limited background knowledge of data set $D$. The attacker just knows a small subset of $D$ and the subset is denoted by $D_{know}$.

\section{membership reference attacks method}

%\par In this section, the main content is about how to perform membership reference attacks against Laplace mechanism. Differential privacy is a widely accepted conception of privacy protection and Laplace mechanism is a famous mechanism to deal with numerical data in the field of differential privacy. Differential privacy is popular because it provides strong privacy guarantee even if an attacker has sufficient background knowledge. However, we find that based on some background knowledge and  linear query, presence or absence of one record can be inferred.% Concretely, first, multiple i.i.d. samples of counting function's answer can be obtained. Second, whether one record is in data set can be judged by hypothesis test based on obtained samples.  

%Because it was based on good mathematics foundation and it provided strong privacy guarantee. It had no assumption about background knowledge of attacker. It also guaranteed that the difference caused by presence or absence of one particular individual's data can not be identical. 
\par In this section, the main content is about how to perform membership reference attacks against Laplace mechanism. An introduction of idea is given before the detailed information of membership reference attacks method is given. Some notations are introduced for simplicity of description. Attackers want to judge whether a record $x$ is in data set D. The background knowledge of attackers is denoted by $D_{know}$ which is a subset of data set $D$, namely $D_{know} \subset D$ and attackers can issue query $q_s(D)$ to Laplace mechanism $M$, denoted by $M(q_s,D,\epsilon)$. The $q_s(D)$ means computing query $q$ over records of data set $D$ under conditions $s$. In other words, the $q_s(D)$ is equal to $q(D\cap D_s)$ where the $D_s$ is the set which satisfies conditions $s$. For example "count the number of students whose age is greater than 10 in classroom". The query $q$ is counting query, the data set $D$ is the set of all the students in classroom and the condition $s$ is that "age is greater than 10".
\par The membership reference attacks method consists of two subroutines. The first subroutine is to obtain multiple i.i.d. samples for linear query's answer. The first subroutine is feasible because of the linear property of linear queries and the detailed reasons are given later. By the first subroutine, attackers can obtain $m$ i.i.d. samples of $M(q_s,D,\epsilon)$ where  $D_s=D_{know} \cup \{x\}$. When the $x$ is in data set $D$, 
\begin{eqnarray*}	
    q_s(D) &=& q(D_s\cap D) \\
    &=& q((D_{know}\cup\{x\})\cap D)\\
    &=& q(D_{know}\cup\{x\}) 
\end{eqnarray*}
%the $q_s(D)$ is equal to $q(D_{know}\cup\{x\})$ 
\par So the mean of output distribution of $M(q_s,D,\epsilon)$ is $q(D_{know}\cup\{x\})$.
\par When the $x$ is not in data set $D$
\begin{eqnarray*}	
    q_s(D) &=& q(D_s\cap D) \\
    &=& q((D_{know}\cup\{x\})\cap D)\\
    &=& q(D_{know}) 
\end{eqnarray*}
%the $q_s(D)$ is equal to $q(D_{know})$ s
\par So the mean of output distribution of $M(q_s,D,\epsilon)$ is $q(D_{know})$. 
\par Attackers can judge whether the $x$ is in the data set $D$ by comparing $q(D_{know})$ and the mean of $M(q_s,D,\epsilon)$. The procedure of comparison is implemented by hypotheses test denoted by $H$ in the second subroutine. 

%\par In the membership reference attacks method, the first subroutine is a  procedure to obtain multiple i.i.d. samples of a linear query under constrains of Laplace mechanism. And the second subroutine is hypothesis test to judge whether $m$ i.i.d. samples come from a distribution with some certain mean. That is, $H$ is to judge whether the samples $\hat{a}_1,\hat{a}_2,\dots,\hat{a}_m$ are drawn from a distribution with mean $q(D_{know})$. When mean of distribution from which $\hat{a}_1,\hat{a}_2,\dots,\hat{a}_m$ are drawn is judged to be $q(D_{know})$, the assertion is that $x$ is not in data set $D$. Otherwise, the assertion is that $x$ is in data set $D$. Next, the detailed information of the two subroutine is given.     

\par In the second subroutine, hypothesis test $H$ is to judge whether the samples $\hat{a}_1,\hat{a}_2,\dots,\hat{a}_m$ are drawn from a distribution with mean $q(D_{know})$. When mean of distribution from which $\hat{a}_1,\hat{a}_2,\dots,\hat{a}_m$ are drawn is judged to be $q(D_{know})$, the assertion is that $x$ is not in data set $D$. Otherwise, the assertion is that $x$ is in data set $D$. Next, the detailed information of the two subroutine is given. 

%When mean of distribution from which $\hat{a}_1,\hat{a}_2,\dots,\hat{a}_m$ are drawn is judged not to be $q(D_{know})$, the assertion is that $x$ is not in data set $D$. Next, the detailed information of the two subroutine is given.     

%a number is equal to mean of distribution by $m$ i.i.d. samples of the distribution. That is, $H$ is to judge whether $q(D_{know})$ is equal to the mean distribution from which the $\hat{a}_1,\hat{a}_2,\dots,\hat{a}_m$ are drawn. When the $q(D_{know})$ is judged to be mean of $\hat{a}_1,\hat{a}_2,\dots,\hat{a}_m$ by $H$, the assertion is that $x$ is not in data set $D$. When the $q(D_{know})$ is not judged to be mean of $\hat{a}_1,\hat{a}_2,\dots,\hat{a}_m$ by $H$, the assertion is that $x$ is in data set $D$. Next, the detailed information of the two subroutine is given. 

\begin{algorithm}
    \caption*{Membership Reference Attacks Method A}
    \label{alg:Framwork} 
    \begin{algorithmic}[1]
        \REQUIRE ~~\\ %算法的输入参数：Input
        the background knowledge of attackers $D_{know}$;\\
        the data set $D$;\\
        a record $x$;\\
        the Laplace mechanism $M$;\\
        a linear query $q$;\\
        total privacy budget $\epsilon_t$;\\
        \ENSURE ~~\\
        the $x$ is in data set $D$ or not;\\
        
        \STATE construct conditions $s$ such that $D_s = D_{know} \cup \{x\}]$; 
        \STATE let $\epsilon = \frac{\epsilon_t}{m}$;		
        \FOR{$i=1$ to $m$}
        \STATE $\hat{a}_i = M(q_s,D,\epsilon)$ by the first subroutine;
        \ENDFOR
        
        \STATE assertion $= H(\hat{a}_1,\hat{a}_2,\dots,\hat{a}_m,q(D_{know}))$ by the second subroutine;
        %\IF{ $f(a)$ was computed and the old result is $a_{old}$}
        %\STATE $a' = a_{old}$
        %\ELSE 		 
        %\STATE $a' = f(a)$ such that $dis(a,a') \le b$
        %\ENDIF  
        %\STATE $\hat{a} = a' + Lap(0,\frac{\Delta D}{\epsilon})$
        \RETURN assertion		  
    \end{algorithmic}
\end{algorithm}

\subsection{Multiple Independent Identical Distribution Samples for Linear Query's Answer } 
\par The goal is to obtain multiple i.i.d. samples for linear query's answers and these samples are used in hypothesis test in the second subroutine. We will introduce used notations first.

\par The $x$ presents a record and we want to confirm whether $x$ is in the data set $D$. Query $q$ is linear query and $m$ i.i.d. samples for $M(q_{s},D,\epsilon)$'s answer are denoted by $\hat{a}_1,\hat{a}_2,\dots,\hat{a}_m$ where $s$ represents query condition such that $D_s =  D_{know}\cup\{x\}$.

\begin{algorithm}
    \caption*{Method to Obtain Multiple i.i.d. Samples of $M(q_{s},D,\epsilon)$}
    \label{alg:Framwork} 
    \begin{algorithmic}[1]
        \REQUIRE ~~\\ %算法的输入参数：Input
        the background knowledge of attackers $D_{know}$;\\
        the data set $D$;\\		 
        the Laplace mechanism $M$;\\
        a linear query $q_s$;\\
        the privacy budget $\epsilon_t$ for query $q_s$;\\
        \ENSURE ~~\\
        the $m$ i.i.d. samples $\hat{a}_1,\hat{a}_2,\dots,\hat{a}_m$ ;\\	
        
        \STATE let $\epsilon = \frac{\epsilon_t}{m}$		
        \FOR{$i=1$ to $m$}
        \STATE choose a new subset $D_i \subset D_{know}$;
        \STATE construct query conditions $s_i$ such that $D_{s_i} = D_i\cup \{x\}$;
        \STATE $a_i = M(q_{s_i},D,\epsilon)$;
        \STATE $\hat{a}_i = a_i + q(D_{know}\setminus D_i)$  	 
        \ENDFOR
        
        \RETURN $\hat{a}_1,\hat{a}_2,\dots,\hat{a}_m$		  
    \end{algorithmic}
\end{algorithm}

\par The attackers randomly choose a data set $D_i$ such that $ D_i \subset D_{know}$, construct query conditions $s_i$ such that $D_{s_i} = D_i\cup \{x\}$ where $D_{s_i}$ is a set whose elements satisfy requirements of conditions $s_i$, submit query $q_{s_i}(D)$ to Laplace mechanism and obtain $a_i = M(q_{s_i},D,\epsilon)$ as return. Let $\hat{a}_i = q(D_{know}\setminus D_i) + a_i$. The data set $D_i$ is a subset of $D_{know}$ and the $D_{know}$ is background knowledge of attackers so that the attacker can compute $q(D_{know}\setminus D_i)$ locally by himself.  The $\hat{a}_i$ is an i.i.d. sample of $M(q_{s},D,\epsilon)$. 
\par Through another subset $D_j$ such that $D_j \subset D_{know}$, attackers can get the second sample of $M(q_s,D,\epsilon)$'s answer. So, the attackers can get $m$ samples $\hat{a}_1,\hat{a}_2,\dots,\hat{a}_m$ for query $M(q_s,D,\epsilon)$ as above.

\par \textbf{Theorem 1}: $\hat{a}_1,\hat{a}_2,\dots,\hat{a}_m$ are samples of $M(q_{s},D,\epsilon)$.
\par \textbf{Proof}: 
\par For $\forall i$, we have    
\begin{eqnarray*}
    \hat{a}_i &=& q(D_{know}\setminus D_i) + a_i \\		
    \hat{a}_i &=& q(D_{know}\setminus D_i) + M(q_{s_i},D,\epsilon) \\
    \hat{a}_i &=& q(D_{know}\setminus D_i) + q_{s_i}(D) + Lap(0,\frac{\Delta D}{\epsilon})\\
\end{eqnarray*}

\par We know that $q_{s_i}(D) = q((D_i\cup \{x\})\cap D)$. The $(D_i\cup \{x\})\cap D $ is a subset of $ D_i\cup \{x\}$. And the $ D_i\cup \{x\}$ and $D_{know}\setminus D_i$ are disjoint. So, the $(D_i\cup \{x\})\cap D $ and $D_{know}\setminus D_i$ are disjoint. In addition, $q$ is linear query. We have  

\begin{eqnarray*}	
    & & q(D_{know}\setminus D_i) + q_{s_i}(D)\\ 
    &=& q(D_{know}\setminus D_i) + q((D_i\cup \{x\})\cap D)   \\
    &=& q(D_u ) \\	
\end{eqnarray*}
\par Here $D_u$ is union set of $D_{know}\setminus D_i$ and $(D_i\cup \{x\})\cap D$.
\begin{eqnarray*}	
    D_u &=& (D_{know}\setminus D_i) \cup ((D_i\cup \{x\})\cap D) \\
    D_u &=& ((D_{know}\setminus D_i)\cap D) \cup ((D_i\cup \{x\})\cap D) \\	
    D_u &=& ((D_{know}\setminus D_i) \cup (D_i\cup \{x\}))\cap D) \\
    D_u &=& (D_{know} \cup  \{x\})\cap D) \\
\end{eqnarray*}
\par So, we have 
\begin{eqnarray*}
    q(D_u ) &=& q((D_{know} \cup  \{x\})\cap D )  \\
    q(D_u ) &=& q_s(D)  \\	
\end{eqnarray*}
\par So, we have
\begin{eqnarray*}
    \hat{a}_i &=& q(D_{know}\setminus D_i) + q_{s_i}(D) + Lap(0,\frac{\Delta D}{\epsilon})\\
    \hat{a}_i &=& q_s(D) + Lap(0,\frac{\Delta D}{\epsilon}) \ \ \ \ \ \ \ \ (1)\\
    \hat{a}_i &=& M(q_s,D,\epsilon)	
\end{eqnarray*} 
$\hfill\Box$ 

\par \textbf{Theorem 2}: $\hat{a}_1,\hat{a}_2,\dots,\hat{a}_m$ are i.i.d. samples of $M(q_{s},D,\epsilon)$.
\par \textbf{Proof}: 
\par According to the equality (1) in proof of theorem 1, for $\hat{a}_i$ we know
\begin{eqnarray*}
    \hat{a}_i &=& q_s(D) + Lap(0,\frac{\Delta D}{\epsilon}) \\
\end{eqnarray*}
\par For all $\hat{a}_1,\hat{a}_2,\dots,\hat{a}_m$, the noise $Lap(0,\frac{\Delta D}{\epsilon})$ are i.i.d. samples of Laplace distribution so $\hat{a}_1,\hat{a}_2,\dots,\hat{a}_m$ are i.i.d. samples.
$\hfill \Box$  

\par \textbf{Theorem 3}: The number of records in $D_{know}$ needs to be greater than $\log_2m$.
\par \textbf{Proof}: 
\par As known before, a subset of $D_{know}$ can generate an i.i.d. sample for query's answer and the number of i.i.d. samples is denoted by $m$. So, the number of records in $D_{know}$ should guarantee that subset of $D_{know}$ is equal to or greater than $m$. The set of subset of $D_{know}$ is called power set denoted by $PA$ and the number of records in $D_{know}$ is denoted by $r$. We have,
\begin{eqnarray*}
    m &\le& |PA| \\
    &\le& 2^{|D_{know}|} \\
    &\le& 2^r 
\end{eqnarray*}
\par Here, the $|\ |$ presents the number of element for a set. We have
\begin{eqnarray*}
    r \ge \log_2m
\end{eqnarray*}
$\hfill\Box$       	 
\par The number $m$ is small so the number $r$ is smaller. As mentioned in the first paragraph of this subsection, the $m$ i.i.d. samples are used in hypothesis test. Usually, the number of samples is small in hypothesis test. For example dozens of numbers are enough. So, the $r$ can be small. For example, when $m=1000$, the $r$ could be 10.  
\par \textbf{Theorem 4}: The totally consumed privacy budget of queries $q_{S_1},q_{S_2},q_{S_3},\dots,q_{S_m}$ is $\epsilon_t$.
\par \textbf{Proof}
\par For $\forall i$, the consumed privacy budget of $M(q_{S_i},D,\epsilon)$ is $\epsilon$ and the $\epsilon = \frac{\epsilon_t}{m}$. According to the Sequential Composition Theorem, the totally consumed privacy budget is equal to    
\begin{eqnarray*}
    \sum_{i=1}^{m}\epsilon = \sum_{i=1}^{m}\frac{\epsilon_t}{m} = \epsilon_t
\end{eqnarray*}
$\hfill\Box$ 
\par In a word, the proposed method in this subsection obtains $m$ i.i.d. samples for a query with privacy budget $\epsilon_t$.
\subsection{Hypothesis Test}

\par In this subsection, the goal is to determine whether a record $x$ is in data set $D$ through hypothesis test. The key  is to find an appropriate statistics. That is, it is key to find an appropriate hypothesis test method. The one-sample t-test is chosen. The $T$ statistics follows $T$ distribution and the $T$ statistics is as follows 
\begin{eqnarray*}
    T &=& \frac{\bar{X} - \mu}{\frac{S}{\sqrt{m-1}}} \\
\end{eqnarray*}

\begin{algorithm}
    \caption*{The One-sample t-test H}
    \label{alg:Framwork} 
    \begin{algorithmic}[1]
        \REQUIRE ~~\\ %算法的输入参数：Input
        the samples $\hat{a}_1,\hat{a}_2,\dots,\hat{a}_m$;\\
        the number of samples $m$;\\		 
        the $q(D_{know})$ ;\\
        the significance level $\alpha=0.05$;\\
        \ENSURE ~~\\
        the $x$ is in data set $D$ or not;\\	
        \STATE make\ hypothesis: \\
        the null hypothesis is $H_0$: $\mu = q(D_{know})$; \\
        the alternative hypothesis is $H_1$: $\mu \ne q(D_{know})$; \\
        \STATE calculate $\bar{X} = \frac{\hat{a}_1+\hat{a}_2+\dots+\hat{a}_m}{m}$;\\
        \STATE calculate $S=\sqrt{\frac{(\hat{a}_1-\bar{X})^2+(\hat{a}_2-\bar{X})^2+\dots+(\hat{a}_m-\bar{X})^2}{m}}$;\\
        \STATE calculate $T = \frac{\bar{X} - \mu}{\frac{S}{\sqrt{m-1}}}$;\\
        \STATE find $p$ according to $T$ and freedom $m-1$;\\
        if $p<\alpha$, the assertion is $x\in$ D; \\
        if $p>\alpha$, the assertion is $x\notin D$ \\ 
        \RETURN assertion		  
    \end{algorithmic}
\end{algorithm}

\par Next, we explain the process of reaching conclusion. Firstly, calculate the $T$ statistics as above. Then according to the freedom $m-1$, distribution of $T$ statistics can be identified. Generally, there is a table related to $T$ distribution. With value of $T$ statistics and freedom $m-1$, the $p$ value can be found in the table which is step 5 of hypothesis test above. If the $p$ value is less than significance level $\alpha$, the alternative hypothesis is accepted, which means that samples $\hat{a}_1,\hat{a}_2,\dots,\hat{a}_m$ don't come from a distribution with mean $\mu = q(D_{know})$. So, $x$ is in data set $D$. Otherwise, the $x$ is not in data set $D$.

\par The significance level $\alpha$ is 0.05. There are two widely accepted choices for significance level, namely $\alpha =0.01$ and $\alpha = 0.05$. We desire that as long as difference is statistically significant, the alternative hypothesis could be accepted. So $\alpha =0.05$ is chosen.

\par The one-sample t-test is the appropriate hypothesis test method because the one-sample t-test is used to judge whether some samples are drawn from a distribution with certain mean. The samples $\hat{a}_1,\hat{a}_2,\dots,\hat{a}_m$ are $m$ independent identical distribution samples of $M(q_s,D,\epsilon)$. The distribution of $M(q_s,D,\epsilon)$ is a Laplace distribution with mean $q_s(D)$. And we know 
\begin{eqnarray*} 
    q_s(D)  =  q(D_s\cap D) = q((D_{know}\cup\{x\})\cap D)  
\end{eqnarray*} 
\par And
\begin{eqnarray*}
    D_{know} \subset D   
\end{eqnarray*}

\par So, if $x$ is in $D$,
\begin{eqnarray*}
    q_s(D) = q(D_{know}\cup\{x\})  
\end{eqnarray*}
\par And if $x$ is not in $D$, 
\begin{eqnarray*}
    q_s(D) = q(D_{know})    
\end{eqnarray*}
\par In other words, when $x$ is not in $D$, the mean of Laplace distribution $M(q_s,D,\epsilon)$ is $q(D_{know})$. When $x$ is in $D$, the mean of Laplace distribution $M(q_s,D,\epsilon)$ is $q(D_{know}\cup\{x\})$. So, if the samples $\hat{a}_1,\hat{a}_2,\dots,\hat{a}_m$ are from distribution with mean $q(D_{know})$,  $x$ is not in $D$. This is the reason why the one-sample t-test is chosen to judge whether samples $\hat{a}_1,\hat{a}_2,\dots,\hat{a}_m$ come from distribution with mean $\mu = q(D_{know})$.

\subsection{Accuracy analysis}

\par The hypothesis test has two types error shown in Table 1.
\begin{table*}[!htbp]
    \centering
    \begin{tabular}{|c|c|c|c|}
        \hline
        \diagbox{assertion}{hypothesis}& $H_0$& $H_1$\\ 
        \hline
        $H_0$& $P\{accetp\ H_0|H_0\ is\ ture\} = 1-\alpha = 0.95 $& $P\{accetp\ H_0|H_1\ is\ ture\} = \delta $\\
        \hline
        $H_1$& $P\{accetp\ H_1|H_0\ is\ ture\} = \alpha = 0.05 $ & $P\{accetp\ H_1|H_1\ is\ ture\} = 1-\delta $\\
        \hline		 
    \end{tabular}
    \caption*{Table 1}
\end{table*}

\par \textbf{Theorem 5} The success rate $R$ of membership reference attacks is approximately equal to $\frac{1}{2}(0.95+\int_{-\infty}^{\frac{S\epsilon_t T_{(m-1,0.05)}}{\sqrt{2}m\sqrt{m-1}}} \frac{1}{2\pi}e^{-\frac{t^2}{2}} dt)$.
\par \textbf{Proof}
\par According the Table 1, we have
\begin{eqnarray*}
    R&=&\frac{(1-\alpha)+(1-\delta)}{(1-\alpha) + \delta + \alpha + (1-\delta)} \\
    &=& \frac{1}{2}(2-\alpha-\delta)\\
    &=& \frac{1}{2}(1.95-\delta)\\
\end{eqnarray*}
\par For the definition of one-sample T test, we have 

\begin{eqnarray*}
    & & \alpha \\
    &=& P\{accetp\ H_1|H_0\ is\ ture\}  \\
    &=& P\{  \frac{\bar{X} - \mu_0}{\frac{S}{\sqrt{m-1}}}<T^*|H_0\ is\ ture \} \\
    &=& 0.05	
\end{eqnarray*}
\par The $\frac{\bar{X} - \mu_0}{\frac{S}{\sqrt{m-1}}}$ approximately obeys $T$ distribution. So, the $T^*\approx T_{(m-1,0.05)}$.

\begin{eqnarray*}
    & & \delta \\
    &=& P\{accetp\ H_0|H_1\ is\ ture\} \\
    &=& P\{  \frac{\bar{X} - \mu_1}{\frac{S}{\sqrt{m-1}}}>T^*|H_1\ is\ ture \} 
\end{eqnarray*}
\par The variance of noise distribution (namely Laplace distribution)  is $\sigma =\sqrt{2} \frac{\Delta D}{\epsilon} = \frac{\sqrt{2} }{\epsilon}$. So, we have
\begin{eqnarray*}
    & & \delta \\	
    &=& P\{  \frac{\bar{X} - \mu_1}{\sigma}>\frac{S\ T^*}{\sigma\sqrt{m-1}}|H_1\ is\ ture \} \\
\end{eqnarray*}
\par The $\frac{\bar{X} - \mu_1}{\sigma}$ approximately obeys $N(0,1)$, So 

\begin{eqnarray*}
    & & P\{  \frac{\bar{X} - \mu_1}{\sigma}>\frac{S\ T^*}{\sigma\sqrt{m-1}}|H_1\ is\ ture \} \\	
    &\approx& \int_{\frac{S\ T^*}{\sigma\sqrt{m-1}}}^{+\infty} \frac{1}{2\pi}e^{-\frac{t^2}{2}} dt \\
    &=& 1- \int_{-\infty}^{\frac{S\ T^*}{\sigma\sqrt{m-1}}} \frac{1}{2\pi}e^{-\frac{t^2}{2}} dt \\
\end{eqnarray*}
\par So, 
\begin{eqnarray*}
    & & R \\
    &\approx&\frac{1}{2}(0.95+\int_{-\infty}^{\frac{S T_{(m-1,0.05)}}{\sigma\sqrt{m-1}}} \frac{1}{2\pi}e^{-\frac{t^2}{2}} dt) \\
    &=& \frac{1}{2}(0.95+\int_{-\infty}^{\frac{S\epsilon T_{(m-1,0.05)}}{\sqrt{2}\sqrt{m-1}}} \frac{1}{2\pi}e^{-\frac{t^2}{2}} dt) \\
    &=& \frac{1}{2}(0.95+\int_{-\infty}^{\frac{S\epsilon_t T_{(m-1,0.05)}}{\sqrt{2}m\sqrt{m-1}}} \frac{1}{2\pi}e^{-\frac{t^2}{2}} dt) \\
\end{eqnarray*}

\section{Security Analysis of Laplace Mechanism and Instance of Attack }
\par In this section, the content is divided into three subsections. In the first subsection, an analysis is given about the security of Laplace mechanism including key idea of Laplace mechanism and background knowledge of attackers. In the second subsection, the reason why counting query is an appropriate liner query to perform membership reference attacks is given. In the last subsection, we discuss cases where proposed method will be influenced and we also discuss how to fix it. 
\subsection {Security Analysis of Laplace Mechanism}
\par The security of Laplace mechanism is based on perturbation. Differential privacy requires that presence or absence of one record cannot be identified through mechanism's output. Laplace mechanism satisfies the requirement by perturbation. Concretely, query's answers are perturbed by noise. The noise is drawn from Laplace distribution whose variance is related to two factors. The first factor is strength of privacy guarantee. The second factor is maximum difference of query's answer and the maximum difference is cased by presence or absence of one record. The maximum difference is captured by conception of sensitivity and the sensitivity is denoted by $\Delta D$. The strength of privacy guarantee is captured by conception of privacy budget and privacy budget is denoted by $\epsilon$. The noise distribution's variance is related to ratio of sensitivity $\Delta D$ to privacy budget $\epsilon$. Specifically, the variance is $2(\frac{\Delta D}{\epsilon})^2$. 
\par The key idea of Laplace mechanism is clear: the difference cased by presence or absence of one record is far less than variance of noise distribution, so it is hard to tell that change of Laplace mechanism's output happens because of noise distribution's randomness or because of presence or absence of a record. The idea can be described intuitively by formula $\Delta D \ll 2(\frac{\Delta D}{\epsilon})^2$. 
If privacy budget is small, the difference cased by presence or absence of one record is much less than variance of noise distribution. The purpose of covering difference cased by presence or absence of one record is achieved. In a word, for Laplace mechanism, difference of query's answer is covered by fluctuation of noise distribution. So, the security foundation of Laplace mechanism can be described by formula $\Delta D \ll 2(\frac{\Delta D}{\epsilon})^2$ intuitively.

\par Next, we discuss background knowledge more carefully. It's hard to answer the question of how much background knowledge is needed. Firstly, the background knowledge is hard to quantify. Because there are so many things which can be background knowledge, such as presence or absence of a record, value of a record and correlation between records. In a word, there are various kinds of background information so  there is no good way to quantify background knowledge. Secondly, what kinds of background knowledge is useful depends on the application. For example, for statistical analysis the statistical characters are important and for interactive query system the value of answer is important.

\par Assumption about the background knowledge in the proposed method is weaker than assumption in differential privacy mechanism. Differential privacy mechanism claimed that even if an attacker knows all records in data set except one, the particular record can be protected. That is, the background knowledge in differential privacy mechanism could be all records except one. In the proposed method, background knowledge is a set of records denoted by $D_{know}$ instead of all records. 

\par Hypothesis test is an appropriate tool to construct membership reference attacks. In Laplace mechanism, the difference cased by presence or absence of one record is covered by the fluctuation of noise. But the hypothesis test is designed to determine that fluctuation is just due to randomness or other reasons. In other words, the hypothesis test can be used to determine whether fluctuation of Laplace mechanism's output is just due to randomness of noise or due to presence or absence of a record.    

\subsection { Instance of Attack Based on Counting Query }

\par Counting query is appropriate query to perform membership reference attacks against Laplace mechanism. On the one hand, sensitivity of counting function is 1 and difference cased by presence or absence of any record is also 1. So, difference cased by presence or absence of any record reaches the maximum. That is, the difference cased by presence or absence of a record is most close to variance of noise distribution. On the other hand, the value range of privacy budget is from 0 to positive infinity. In most cases, it is secure if the value of privacy budget is from 0 to 1. Let's check the inequality $\Delta D \ll 2(\frac{\Delta D}{\epsilon})^2$. For counting query, when privacy budget is 1, the sensitivity $\Delta D$ is 1 and $2(\frac{\Delta D}{\epsilon})^2$ is 2. The $\Delta D$ is half of $2(\frac{\Delta D}{\epsilon})^2$. When budget is 0.5, the sensitivity $\Delta D$ is 1 and $2(\frac{\Delta D}{\epsilon})^2$ is 8. So $\Delta D$ is one eighth of $2(\frac{\Delta D}{\epsilon})^2$. In a word, the inequality $\Delta D \ll 2(\frac{\Delta D}{\epsilon})^2$ does not hold for counting query when privacy budget takes values which is generally considered to be an appropriate value for privacy budget.

\par There are other three reasons why query $q$ should be counting query. Firstly, counting query makes the method technically right. Counting query satisfies linear property, $q(D_1 \cup D_2) = q(D_1) + q(D_2)$ and the sensitivity of counting query is always 1. These two properties make it possible to obtain $m$ i.i.d. samples for counting query's answer. In addition, the hypothesis test works on i.i.d. samples. So, the method is technically right because of unique properties of counting query.

\par Secondly, the proposed method is based on counting query so the proposed method can work in a wide range. Counting query is very common. On the one hand, it is a common need to confirm number of records which meet specific requirements, such as $sql$ statement " select count(*) from table student where student.age $\le$ 20 ". On the other hand, there are a lot of complex queries based on counting query. 

\par Thirdly, difference cased by presence or absence of one record is most easily detected by hypothesis test when the query is counting query. As mentioned before, sensitivity $\Delta D$ is maximum difference cased by presence or absence of one record. In addition, noise in Laplace mechanism is drawn from Laplace distribution with variance $2(\frac{\Delta D}{\epsilon})^2$. The difference cased by presence or absence of one record is smaller than sensitivity and the sensitivity is far smaller than variance of noise distribution. So, it is hard to detect the difference cased by presence or absence of one record. When the query is counting query, the difference cased by presence or absence of any record is 1, reaching maximum value, namely sensitivity. So, counting query makes it easy to detect the difference cased by presence or absence of one record.    

\subsection { Discussion}

\par Usually, multiple i.i.d. samples for a query's answer cannot be obtained directly under constrains of Laplace mechanism. Laplace mechanism is a random mechanism so it can output different answers for the same query every time the same query is submitted. However, in order to reduce consumption of privacy budget, it outputs the same answer for queries which have been answered before. To that end, a table is maintained. The table records queries which are answered and answers of these queries. When a query is submitted, the query is searched in the table. If there is answer for the query in the table, the answer will be returned. But, our proposed method to get multiple i.i.d. samples can work well when Laplace mechanism reduces consumption of privacy budget as above. The reason is that proposed method issues different queries every times.  

\par The dependence detection of queries has an influence on proposed method to get multiple i.i.d. samples. In order to strengthen security of differential privacy mechanism, the dependence of queries will be detected by algorithms. That is, when a query comes, algorithms will check whether the query is dependent of queries which have been answered. If there is dependence, the privacy budget will be doubled. 

\par About the influence of dependence detection, there are two claims to clarify. First, there are no effective methods to perform dependence detection. There are so many queries that detecting dependence among these queries is feasible in terms of computing power. Especially, there are various kinds of queries. Second, even if there are effective methods to perform dependence detection, the proposed method can work well after it is improved. Concretely, disjoint subsets of $D_{know}$ can be used to obtain i.i.d. samples, namely $D_i \cap D_j = \varnothing$ for $0<i\ne j<m$ and $D_{know}=D_1\cup D_2 \cup \dots \cup D_m$. When $D_i$ and $D_j$ are disjoint, the query $M(q_{s_i},D)$ and query $M(q_{s_j},D)$ are independent. So, the improved method can obtain multiple i.i.d. samples under constrain of dependence detection. 

\par The improved method requires attackers with more background knowledge. In the proposed method, the number of records in $D_{know}$ is $log_2m$. But in the improved method, the number of records in $D_{know}$ is $\sum_{k=1}^{m}|D_k|$. We know that $\sum_{k=1}^{m}|D_k| \ge m>log_2m$. The equality holds when there is only one record in every $D_k$.    

\par The smaller the sensitivity of query is, the weaker privacy guarantee of Laplace mechanism is. When sensitivity of query is small, the utility(accuracy) will be good under the constrains of Laplace mechanism. So small sensitivity is good property of query in terms of utility or accuracy. But security of query with small sensitivity is vulnerable. We use a formula to describe security foundation of Laplace mechanism intuitively and the formula is $\Delta D \ll 2(\frac{\Delta D}{\epsilon})^2$, namely $\epsilon^2 \ll 2\Delta D$. When sensitivity $\Delta D$ is small such as 0.5, the formula is not easy to hold. We will give an experiment to demonstrate the phenomenon.

\section{Experiments}
\par To demonstrate the efficiency of our algorithm, we comprehensively compare our algorithm with the state-of-the-art and classical algorithms through extensive experiments.

\par The first experiment focuses on the influences of smoothness bound on the time complexity of our algorithm and the index calculus algorithm. In the index calculus algorithm, smoothness bound $B$ significantly influences its computing speed because the smoothness bound $B$ determines the size of factor base. As smoothness bound $B$ increases, the factor base becomes larger, and the computation speed of index calculus algorithm increases first and then decreases. In previous work of Andrew \cite{add_1},  it is proved that the optimal value of $B$ is 	
\begin{equation}
    e^{\sqrt{\frac{\log{p} \log{\log{p}}} {2} }}, \label{equ_22} 
\end{equation}
here $p$ is the order of finite prime field. That is, the index calculus algorithm achieves its maximal computation speed when $B$ is set according to (\ref{equ_22}). 
\par To explore the influences of smoothness bound on time complexity of algorithms, many random discrete logarithm problems are generated and then different smoothness bound is used to solve these discrete logarithm problems. In particular, the bits length of order of finite prime field is from 30 bits to 50 bits with step size 2 and the smoothness bound is set  $ \{ 0.1B_i, 0.5B_i, 1B_i, 1.5B_i, 2B_i \}$ where 
\begin{eqnarray}
    B_i = e^{\sqrt{\frac{\log{p} \log{\log{p}}} {2} }},
\end{eqnarray}
and $p$ is the order of finite prime field. The average running time of solving these discrete logarithm problems are calculated and shown in Table \ref{table_all}.

\par
\begin{figure}[htbp]
    \centering  
    \includegraphics[scale=0.5]{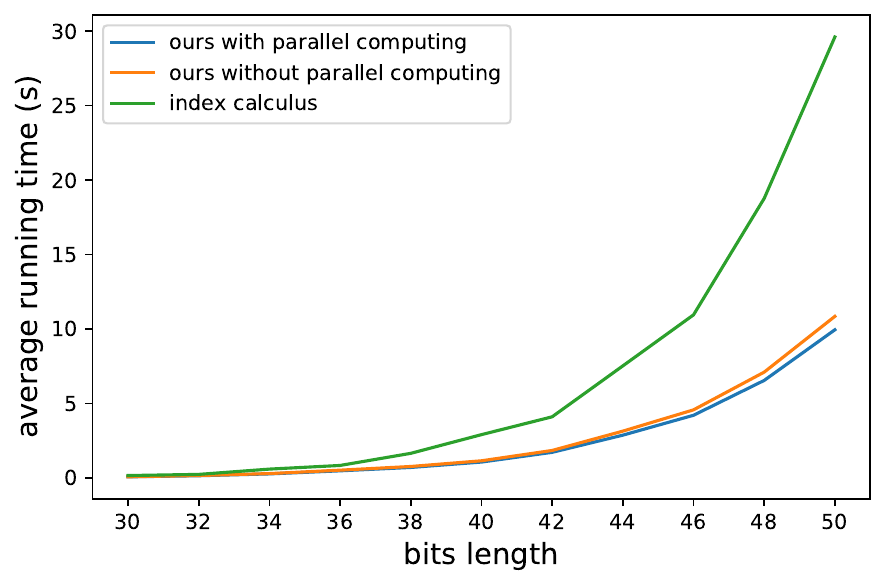}
    \caption{ Comparison of Average Running Time of Smoothness Bound $0.5e^{\sqrt{{\log{p} \log{\log{p}}}/{2} }}$}
    \label{smooth bound}
\end{figure}

 According to the experiment results in Table \ref{table_all}, the average running time increases as the bits length increases in both our double index calculus algorithm and the index calculus algorithm. The experiment results also indicate that the running time decreases first and then increases as the smoothness bound increases for each bits length. When the smoothness bound is \begin{equation}
    0.5e^{\sqrt{\frac{\log{p} \log{\log{p}}} {2} }},  
\end{equation}
the average running time of all algorithms achieves the minimal value. To demonstrate the comparison among algorithms more clearly, the average running time of all algorithms with smoothness bound $0.5e^{\sqrt{\frac{\log{p} \log{\log{p}}} {2} }}$ is shown in Figure \ref{smooth bound}.

\par According to the experiment results in Figure \ref{smooth bound}, our algorithm is much fast than the index calculus algorithm. In particular, our algorithm is about three times ($29.6083/9.9497 \approx 3$) faster than the index calculus algorithm. 

\par
\begin{figure}[htbp]
    \centering  
    \includegraphics[scale=0.5]{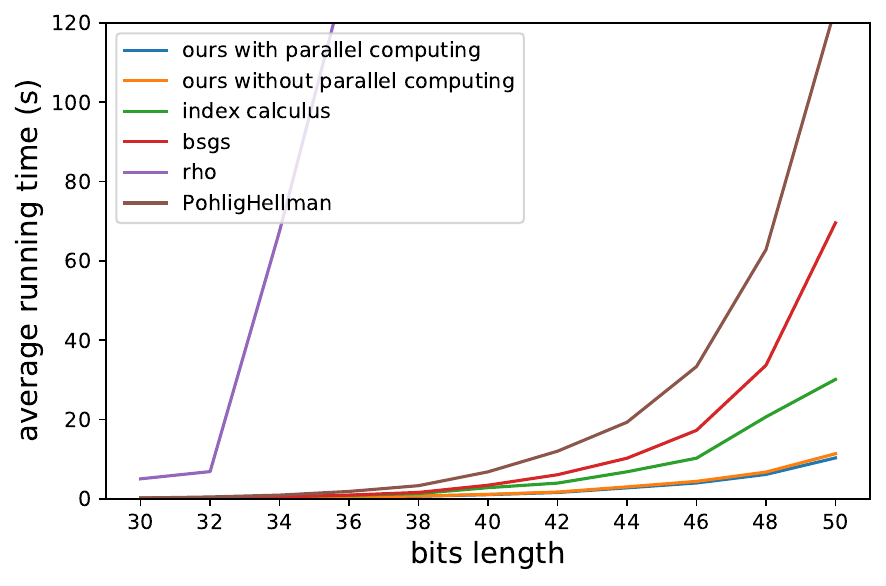}
    \caption{ Comparison of Average Running Time}
    \label{all_algorithms}
\end{figure}

 In the second experiment, we compare our algorithm with other classic and state-of-the-art algorithms, including index calculus algorithm, baby step giant step algorithm,  PohligHellman algorithm and Rho algorithm. The bits length of order of the finite prime field is from 30 bits to 50 bits with step size of 2. The smoothness bound is $0.5e^{\sqrt{\frac{\log{p} \log{\log{p}}} {2} }}$. Many random discrete logarithm problems are generated and the average running times of solving these discrete logarithm problems are shown in Figure \ref{all_algorithms}. 
\par The experiment results in Figure \ref{all_algorithms} indicate that our algorithm is the most efficient algorithm to solve discrete logarithm problems. In particular, the running time of index calculus algorithm, baby step giant step algorithm, and PohligHellman algorithm are about 3 times, 7 times and 12 times that of our algorithms. The average running time of rho algorithm is far more than 20 times that of our algorithm.   

\par
\begin{figure*}[htbp] %通栏
    \begin{minipage}[t]{1.0\linewidth} %调节两个子图左右间距
        \centering
        \includegraphics[width=5.6cm,keepaspectratio]{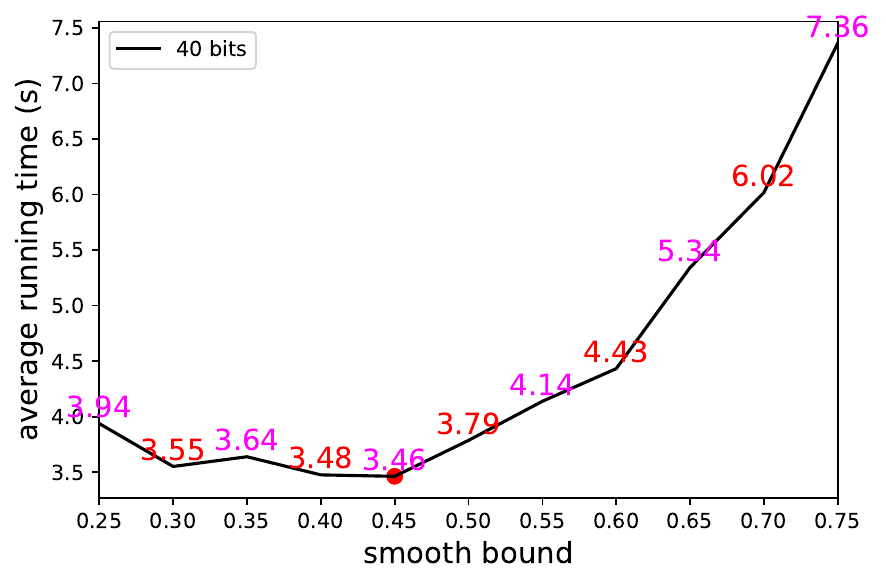}
        \includegraphics[width=5.6cm,keepaspectratio]{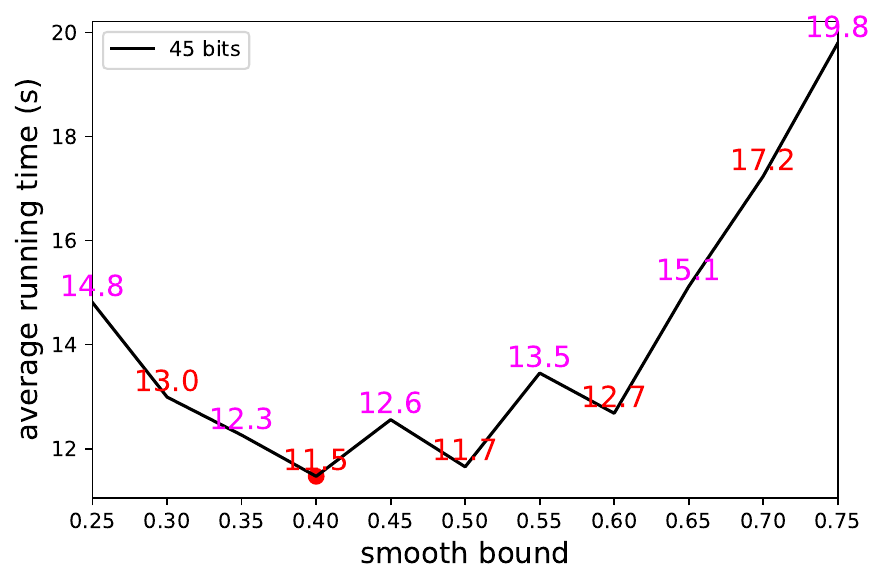}
        
        \includegraphics[width=5.6cm,keepaspectratio]{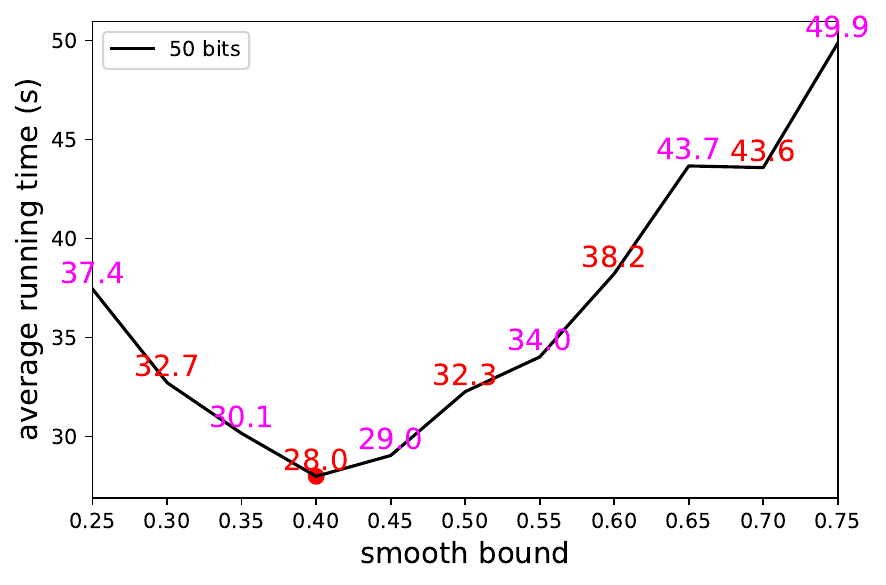}
        \includegraphics[width=5.6cm,keepaspectratio]{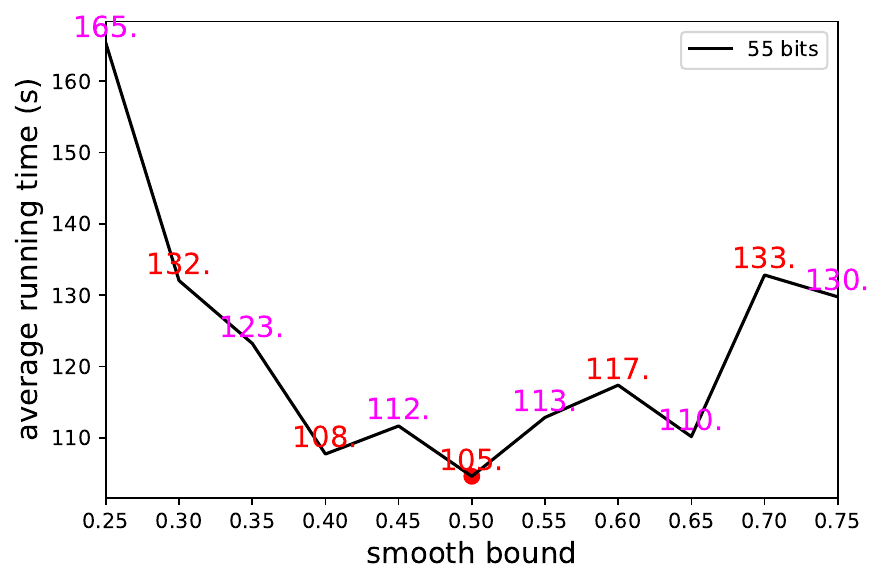}  
        \centerline{(a) Index Calculus Algorithm}
        %\caption{Index Calculus Algorithm} %子图下标题
        %\label{fig1} %引用标签
    \end{minipage}%

    \begin{minipage}[t]{1.0\linewidth} %调节两个子图左右间距
        \centering
        \includegraphics[width=5.6cm,keepaspectratio]{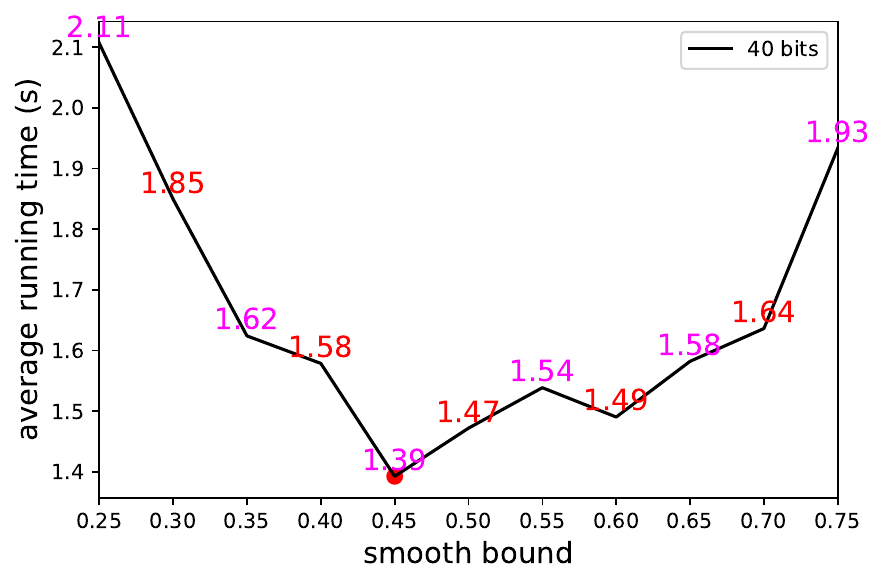}
        \includegraphics[width=5.6cm,keepaspectratio]{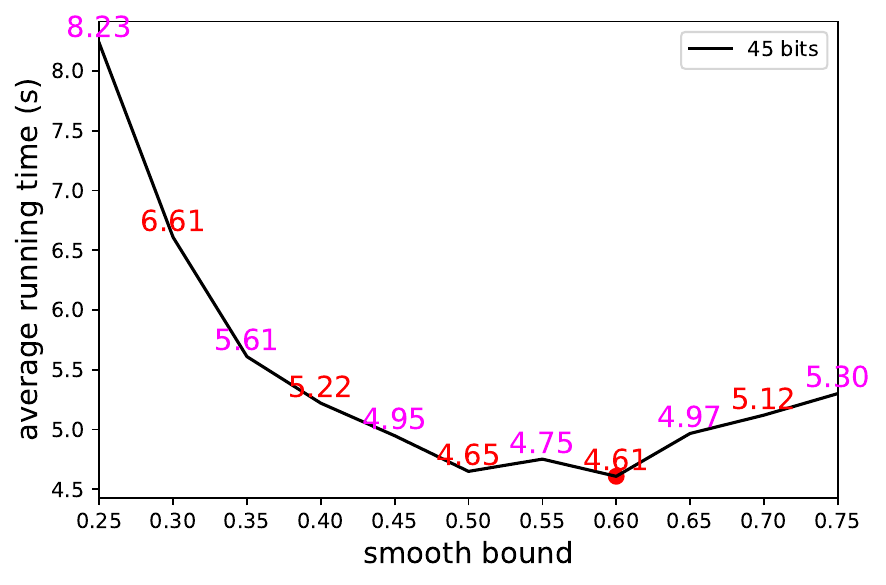}
        
        \includegraphics[width=5.6cm,keepaspectratio]{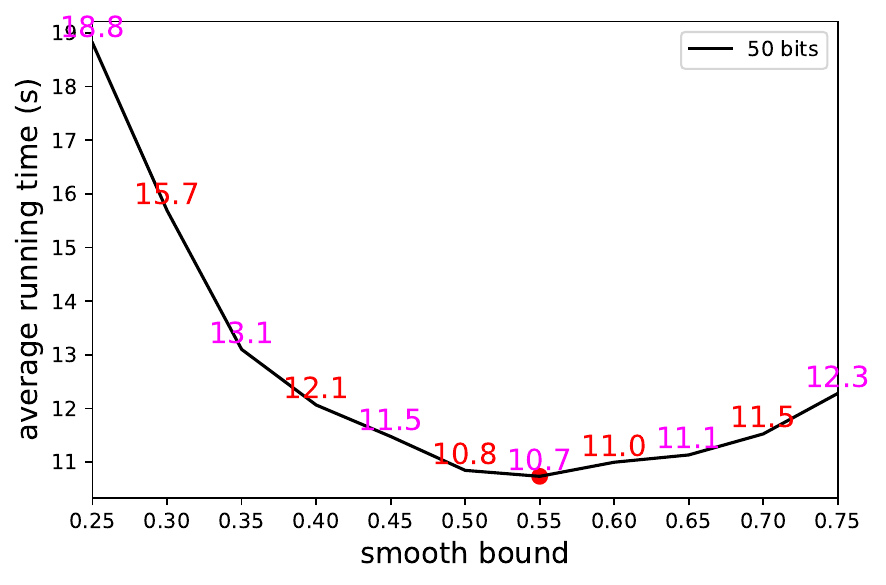}
        \includegraphics[width=5.6cm,keepaspectratio]{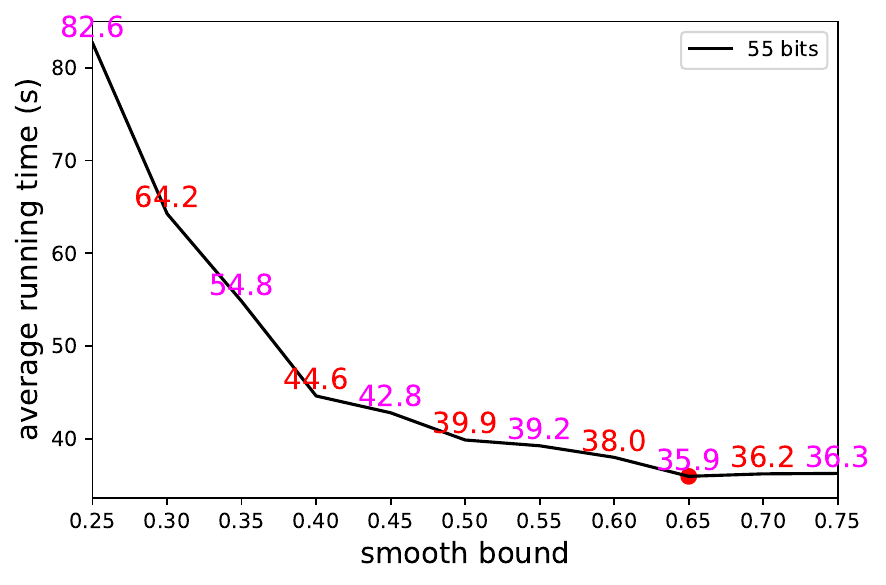}  
        \centerline{(b) Ours without Parallel Computing}
        %\caption{Ours without Parallel Computing} %子图下标题
        %\label{fig2} %引用标签
    \end{minipage}%

    \begin{minipage}[t]{1.0\linewidth} %调节两个子图左右间距
        \centering
        \includegraphics[width=5.6cm,keepaspectratio]{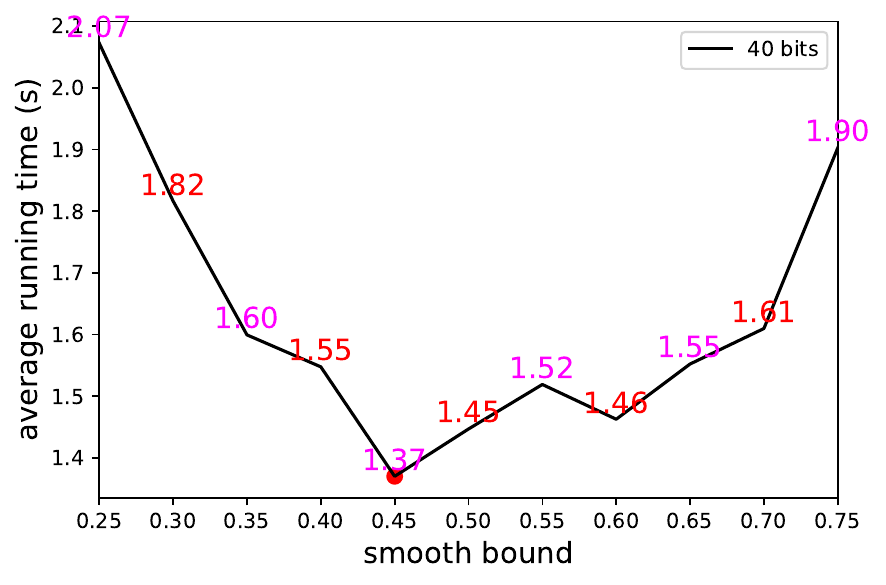}
        \includegraphics[width=5.6cm,keepaspectratio]{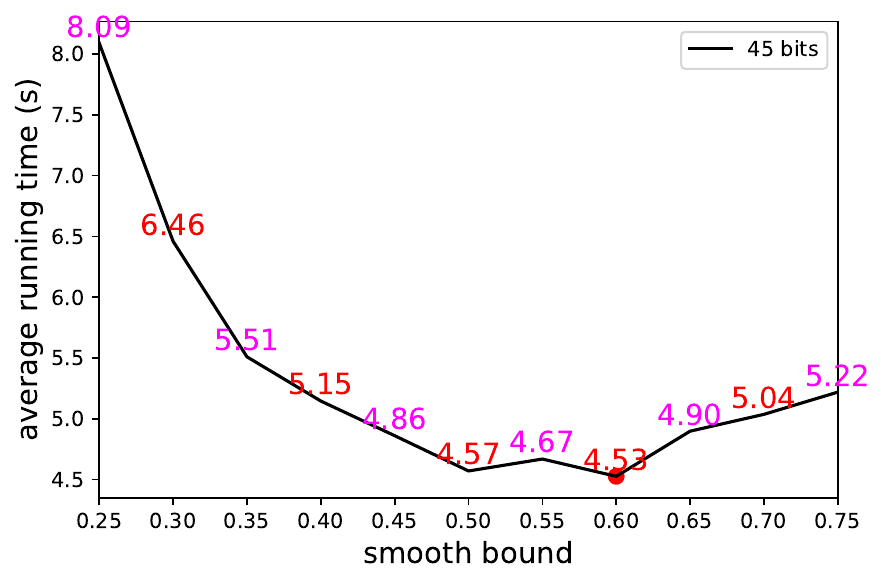}
        
        \includegraphics[width=5.6cm,keepaspectratio]{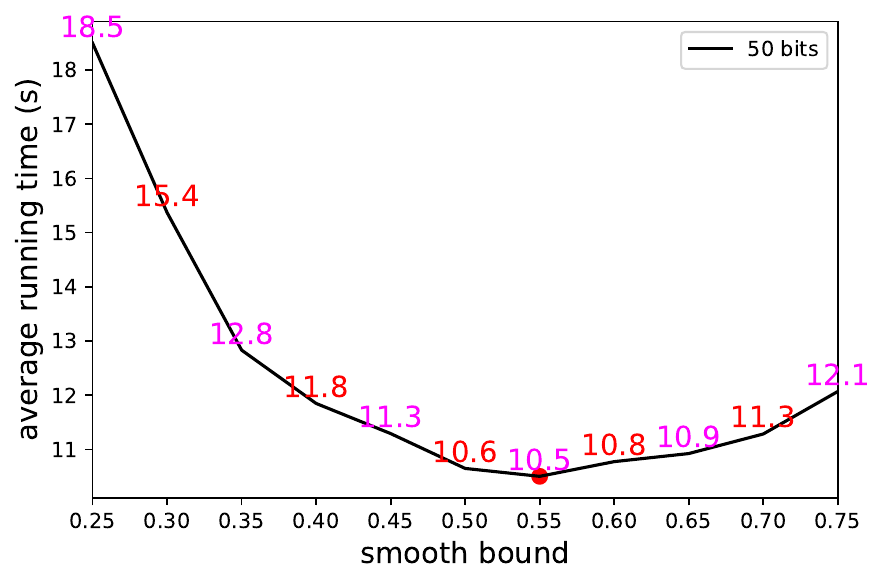}
        \includegraphics[width=5.6cm,keepaspectratio]{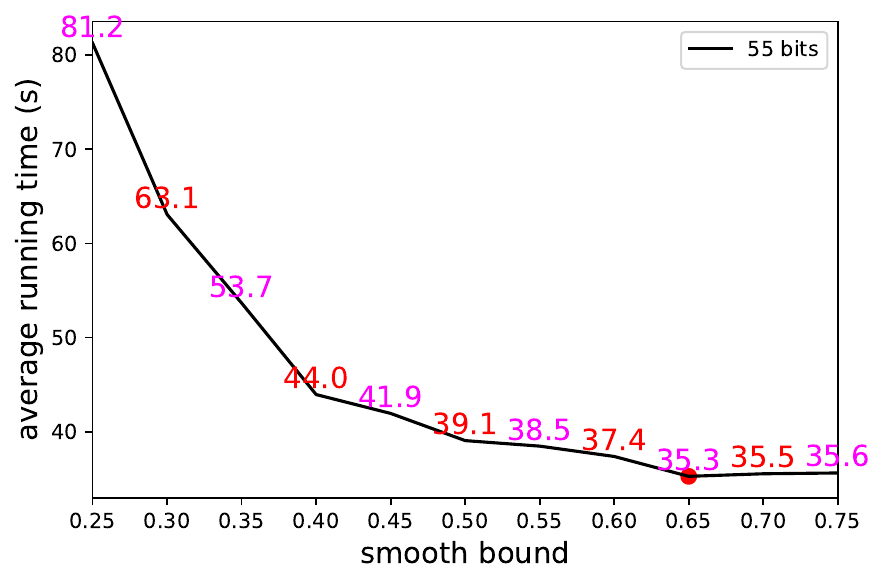}  
        \centerline{(c) Ours with Parallel Computing}
        %\caption*{Ours with Parallel Computing} %子图下标题
        \label{fig3} %引用标签
    \end{minipage}%
    \caption{Influences of Smooth Bound on Average Running Time. The minimum of average running time is marked with a solid red dot.}
    \label{minimum}
\end{figure*}

 The third experiment focuses on the minimum of average running time. According to the experiment results in the first experiment, the average running time may be minimal when the smoothness bound is $0.5e^{\sqrt{\frac{\log{p} \log{\log{p}}} {2} }}$. To explore the minimum of the average running time, the smooth bounds are set to be {$0.25B_i$, $0.30B_i$, $0.35B_i$, $0.40B_i$, $0.45B_i$, $0.50B_i$, $0.55B_i$, $0.60B_i$, $0.65B_i$, $0.70B_i$}, where       $B_i = e^{\sqrt{\frac{\log{p} \log{\log{p}}} {2} }}$. The bit length is from 40 to 55 with step size 5. Many random discrete logarithm problems are generated and average running time of solving these discrete logarithm problems are shown in Fig \ref{minimum}.

\par The experiment results in Fig. \ref{minimum} demonstrate obvious trends. First, the average running time decreases and then increases after a specific value as the smoothness bound increases from $0.25B_i$ to $0.75B_i$, where $B_i=e^{\sqrt{\frac{\log{p} \log{\log{p}}} {2} }}$. For example, the average running time of index calculus algorithm decreases when smoothness bound is  smaller than $0.45B_i$. And then, the average running time of index calculus algorithm increases when smoothness bound is  larger than $0.45B_i$. Second, our algorithm is faster than index calculus algorithm. In particular, when solving random discrete logarithm problems, the minimal average running time of index calculus algorithm is three times that of our algorithm. For example, the minimal average running time of index calculus algorithm is about 105s when the bit length is 55 bits. The minimal average running time of our algorithm is only about 1/3 (105/35.3) that of index calculus algorithm when the bit length is 55 bits. Third, the value of smoothness bound that makes the average running time minimal is become larger as the bit length increase. For example, the average running time is minimal of all algorithms when smoothness bounds is $0.45B_i$ if the bit length is 40. However, the average running time is minimal of our algorithm and index calculus algorithm when smoothness bounds is $0.65B_i$ and $0.50B_i$ respectively if the bit length is 55.

\par
\begin{figure}[htbp]
    \centering  
    \includegraphics[scale=0.5]{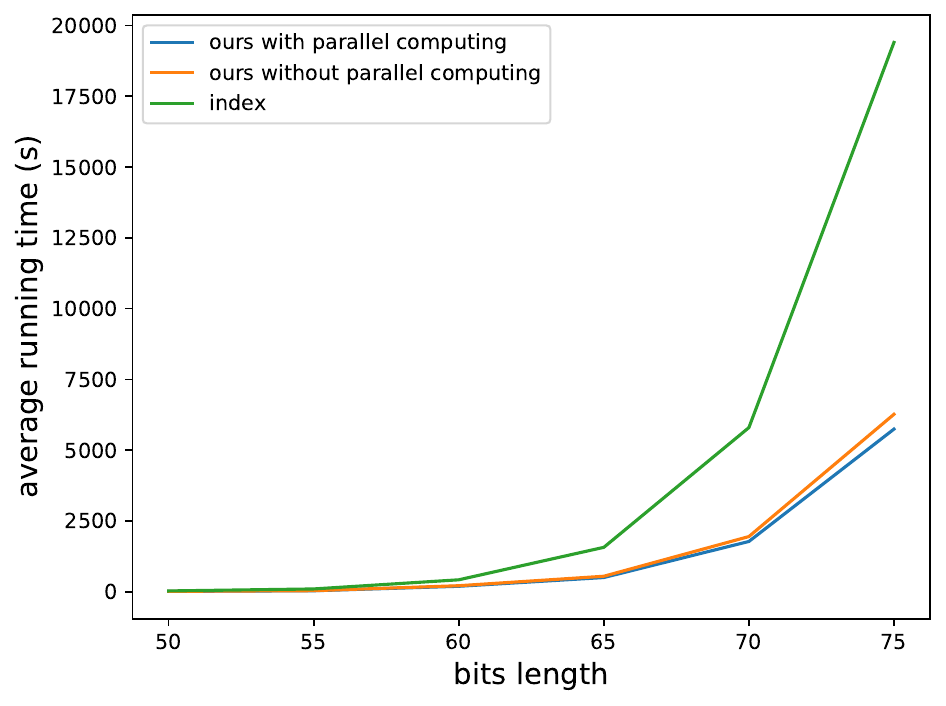}
    \caption{ Comparison of Average Running Time for Large Bit Length}
    \label{high_bits}
\end{figure}

 The fourth experiment compares the average running time when the bit length is large. In particular, bit length is from 50 bits to 75 bits with step size 5. As demonstrated in the third experiment, the smoothness bound has large influence on the average running time. For the index calculus algorithm, the average running time is minimal when smoothness bound is $0.5B_i$, where $B_i=e^{\sqrt{\frac{\log{p} \log{\log{p}}} {2} }}$. And for our algorithm, the average running time is minimal when smoothness bound is $0.65B_i$. Thus, in the fourth experiment, smoothness bounds of index calculus algorithm and our algorithm are $0.5B_i$ and $B_i$, respectively. Many random discrete logarithm problems are generated and the average running time of solving these problems is shown in Fig. \ref{high_bits}.
   
\par According to the experiment results in Fig. \ref{high_bits}, the computing speed of our algorithm is higher than that of index calculus algorithm. As the bit length becomes large, the difference on average running time becomes large too, which indicates that the performance of our algorithm becomes more better than the performance of the state-of-the-art index calculus algorithm.

\par To demonstrate that our algorithm could be much faster than the state-of-the-art index calculus algorithm, we randomly set smooth bound to be \begin{eqnarray}
    B_i = e^{\frac{\sqrt{\log p\log\log p}}{2}} \label{euq_72}
\end{eqnarray}   
and 
\begin{eqnarray}
    B_d = e^{\sqrt{\frac{\log p\log\log p}{2}}} \label{euq_73}
\end{eqnarray}  
for index calculus algorithm and our algorithm. The comparison of average running time is shown in Table \ref{table-5} and the comparison indicates that our algorithm could be 30 times faster than the state-of-the-art index calculus algorithm.

\begin{table}[h]
    \centering
    \caption{Comparison of the Average of Running Time }
    \label{table-5}
    \begin{tabular}{|c|c|c|c|c|c|}
        \hline
        \multirow{2}{*}{}  & \multicolumn{5}{c|}{ bit length }\\
        \cline{2-6} 
        & 30 bits & 40 bits & 50 bits & 60 bits & 70 bits \\
        \hline 
        \makecell{index\\ calculus\\ algorithm} & 0.3s & 5.36s & 114.77s & 1707.66s & 75426.04s \\
        \hline
        \makecell{our\\ algorithm }& 0.15s &1.32s &11.67s&163.46s & 2211.92s \\
        \hline
         times & 2  & 4.1 & 9.8& 10.4 & 34.1 \\
        \hline 		
    \end{tabular}
\end{table}

\par In brief, our algorithm is compared with classical and state-of-the-art algorithms comprehensively, and the comparison indicates that our algorithm is better than existing algorithms.   

\balance
\section{Conclusion}

\par In this paper, we propose the double index calculus algorithm to solve the discrete logarithm problem in the finite prime field. Our algorithm is faster than the fastest algorithm available. We give theoretical analyses and perform experiments to back up our claims. Empirical experiment results indicate that our algorithm could be more than 30 times faster than the fastest algorithm available when the bit length of the size of the prime field is large. Our improvement in solving the discrete logarithm problem in the finite prime field may influence how researchers choose security parameters of cryptography schemes whose security is based on the discrete logarithm problem.

\bibliographystyle{IEEEtran}
\bibliography{mybibliography}

\end{document}